\documentclass[11pt]{article}
\usepackage{fullpage}
\usepackage[utf8]{inputenc}
\usepackage[english]{babel}
\usepackage[T1]{fontenc}
\usepackage{amsmath,amssymb,amsfonts,amsthm}
\usepackage{url}
\usepackage{hyperref}
\usepackage{cleveref}
\usepackage{graphicx}
\usepackage[ruled,vlined, linesnumbered]{algorithm2e}

\newtheorem*{theorem*}{Theorem}

\usepackage{authblk}

\newcommand{\ket}[1]{|{#1}\rangle}

\newcommand{\Prob}{\mathbb{P}}

\DeclareMathOperator{\fail}{fail}
\DeclareMathOperator{\dec}{dec}
\newcommand{\tmax}{t_{\max}}

\DeclareMathOperator{\SEC}{SEC}
\newcommand{\tSEC}{t_{\SEC}}
\DeclareMathOperator{\range}{range}
\DeclareMathOperator{\cost}{cost}
\DeclareMathOperator{\mincost}{mincost}
\DeclareMathOperator{\argmincost}{argmincost}

\usepackage{xcolor}

\title{How to choose a decoder for a fault-tolerant quantum computer? 
The speed vs accuracy trade-off
}

\author[]{Nicolas Delfosse, Andres Paz, Alexander Vaschillo, Krysta M.~Svore}

\affil[]{Microsoft Quantum, Redmond, Washington 98052, USA}

\begin{document}

\maketitle

\date

\begin{abstract}
Achieving practical quantum advantage requires a fault-tolerant quantum computer, namely a quantum machine that operates reliably over so-called logical qubits and logical operations.  Such fault-tolerant logical computation necessarily relies not only on quantum error correction, but also on a classical decoding algorithm to identify and correct faults during computation. This classical decoding algorithm must deliver both accuracy and speed, but in what combination?  When is a decoder "fast enough" or "accurate enough"?

In the case of surface codes, tens of decoding algorithms have been proposed, with different accuracies and speeds. However, it has been unclear how to choose the best decoder for a given quantum architecture. Should a faster decoder be used at the price of reduced accuracy? Or should a decoder sacrifice accuracy to fit within a given time constraint?  If a decoder is too slow, it may be stopped upon reaching a time bound, at the price of some time-out failures and an increased failure rate. What then is the optimal stopping time of the decoder?

By analyzing the speed vs.~accuracy tradeoff, we propose strategies to select the optimal stopping time for a decoder for different tasks. We design a protocol to select the decoder that minimizes the spacetime cost per logical gate, for logical computation of a given depth. 
Our protocol enables comparison of different decoders, and the selection of an appropriate decoder for a given fault-tolerant quantum computing architecture. 
We illustrate our protocol for the surface code equipped with a desktop implementation of the PyMatching decoder. 
We estimate PyMatching is fast enough to implement thousands of logical gates with a better accuracy than physical qubits.
However, we find it is not sufficiently fast to reach $10^5$ logical gates, under certain assumptions, due to the decoding delay which forces qubits to idle and  accumulate errors while idling.
We expect further improvements to PyMatching are possible by running it on a better machine or by reducing the OS interference.
\end{abstract}

\section{Introduction}
\label{sec:introduction}

Achieving {\it practical quantum advantage} \cite{hoefler2023disentangling} will require achieving each of three quantum computing implementation levels (QCILs) \cite{msftblog}, and ultimately demonstrating the third QCIL, {\it scale}.
Today, several types of quantum computers are able to achieve the first QCIL, {\it foundational}, and are referred to as "Noisy Intermediate Scale Quantum" (NISQ) machines~\cite{preskill2018quantum}.
However, demonstrating the ability to outperform a classical machine for a useful problem in a reasonable amount of time requires reaching scale, namely a quantum supercomputer able to execute $10^8$ logical $T$ gates reliably.

Before reaching a scaled quantum supercomputer, quantum computers will first need to demonstrate the second QCIL, {\it resilient}, in which the machine  will showcase fault-tolerant operations over logical qubits such that the logical computation is able to outperform the corresponding physical computation.  Ultimately, the goal at this QCIL is to demonstrate the ability to extend the depth of reliable computation, translating to enabling lower and lower logical error rates across an increasing number of logical qubits.

To enable scaling up fault-tolerant computation, both a quantum error correction code and a corresponding classical decoding algorithm, called a decoder, are required.  The decoder and its performance are critical; it is a classical algorithm whose role is to identify the effect of faults occurring during the computation based on so-called syndrome measurement data, so that these faults can be corrected before they spread throughout the quantum computer.
Despite extensive progress in the field of quantum error correction and decoding, a crucial question remains open:

\bigskip
{\it 
Q1. How to choose a decoder for a fault-tolerant quantum computer?
}
\bigskip

To answer question Q1, we provide a general procedure to select an appropriate decoder for a fault-tolerant quantum computing architecture.
Selecting a decoder, naturally, depends on several  constraints introduced by the underlying hardware architecture.
Ref.~\cite{beverland2022assessing} assesses the requirements for scaling a hardware architecture, concluding that requirements of a scalable qubit design are fast speed, reliable control, and small size. Two of these requirements, namely the physical qubit operation speed and reliability, are critical inputs to consider when selecting a decoder.
For example, the constraint on the speed of a decoder is relaxed if physical qubit operations are slower.

In this work, we illustrate our approach with the surface code architecture~\cite{dennis2002topological, raussendorf2007fault, fowler2012surface} because it is simple and well understood.\footnote{We refer the reader to Ref.~\cite{litinski2019game} for an overview of fault-tolerant quantum computing with the surface code.}
However, our procedure immediately generalizes to other codes as well.
We consider a physical error rate of $p=10^{-3}$, and assume a specific set of fault-tolerant logical operations, using a specific algorithm for compilation of logical gates. 
We assume one round of syndrome extraction may be implemented in 1$\mu s$.
This setting is reasonably well-suited to a variety of  hardware architectures with fast physical operation speeds, such as Majorana~\cite{kitaev2001unpaired, sarma2015majorana, karzig2017scalable}, spin~\cite{kane1998silicon, hanson2007spins, jnane2022multicore}, and superconducting based qubit architectures~\cite{arute2019quantum, hong2020demonstration,kjaergaard2020superconducting, koch2007charge, schreier2008suppressing, steffen2011quantum}.

While several quantum computing hardware architectures have gotten closer to realizing qualities of the resilient QCIL 
~\cite{zhao2022realization, krinner2022realizing, bluvstein2022quantum, google2023suppressing},
the limitation in these experiments is {\it not} the decoding algorithm.  Rather, these experiments rely on small decoders, and any small code decoder can be implemented with a lookup table storing all possible corrections.
Moreover, for a Clifford-based computation or memory experiment, decoding can be done entirely offline, and  any decoding delay does not affect the experiment.

In this work we focus on the requirements of a decoder for realizing a quantum supercomputer at the scale QCIL.
It is motivated by the regime of practical applications which typically require upwards of a thousand logical qubits, encoded in millions of physical qubits~\cite{reiher2017elucidating, campbell2019applying, gidney2021factor}.
Decoding millions of qubits simultaneously presents significant challenges for a decoder design. 
The decoder must be fast enough to avoid the accumulation of errors and capable of correcting a thousand or more logical qubits simultaneously.

An efficient decoder for the surface code based on a Minimum Weight Perfect Matching (MWPM) algorithm was proposed in 2001~\cite{dennis2002topological} and was carefully optimized over the past 20 years, in particular by Fowler~\cite{fowler2012towards}, and most recently Higgott~\cite{higgott2021pymatching} and Higgott and Gidney~\cite{higgott2023sparse} with the PyMatching implementation~\cite{pymatching2023}. 
Another decoding strategy is the Union-Find (UF) decoder~\cite{delfosse2021almost} which has a more favorable worst-case complexity, but which is less accurate.
Many other surface code decoders were proposed; see Refs.~\cite{battistel2023real} or~\cite{iolius2023decoding} for recent reviews.
Recent works argue that decoders, such as the UF decoder running on specialized hardware, can be made fast enough for large-scale fault-tolerant quantum computing~\cite{das2022afs, holmes2020nisq+, ueno2021qecool}.
Variants of the UF decoder implemented on an FPGA show promising performance and may already be fast enough to decode surface codes~\cite{liyanage2023scalable, barber2023real}.
Given that decoders exist that show encouraging accuracy and performance, it is important to answer question Q1, how to choose a decoder, by designing a protocol to select the most adapted decoder for a given fault-tolerant quantum computing architecture.

Decoders are typically compared with one another by considering their accuracy, {\em i.e.}, the decoding failure rate they achieve, or their average or worst-case complexities.
Focusing on the decoder accuracy is not enough in general because the most accurate decoder, that is the maximum likelihood decoder, is often too slow in practice.
Average or worst-case complexities are insightful indicators for the speed of a decoder, but they do not provide enough information to choose between two decoders. 
In practice, the exact runtime matters.
In this work, we provide an answer to Q1 by analyzing the speed vs.~accuracy tradeoff for decoders.

First, consider a simpler question. Given a decoder and a fault-tolerant quantum computer architecture, is there a maximum  runtime that should be assigned to the decoder, beyond which the decoder becomes useless or even harmful? Or in other words:

\bigskip
{\it 
Q2. Can the decoder be too slow?
}
\bigskip

If we only consider Clifford-based computation, decoding can be performed entirely offline, after the computation is complete. 
In this case, there is no strict theoretical limit on the decoding time, even though a short post-processing time is more convenient. 
Moreover, this decoding work can be parallelized to reduce the post-processing time~\cite{skoric2022parallel, tan2022scalable, bombin2023modular}.

In contrast, in the case of a universal quantum computer, where non-Clifford operations are invoked, the decoding problem comes with different requirements. 
For example, logical $T$ operations are required.
The implementation of a logical $T$ gate leads to the introduction of idle steps due to a decoding delay.
Fig.~\ref{fig:Tgate} shows the standard $T$ state injection circuit, which includes a classically-controlled $S$ gate conditioned on the outcome of the measurement of a logical ancilla qubit. 
Because this measurement outcome is extracted by the decoder, one must wait for the decoder to terminate to know if the $S$ gate has to be performed or not.
Terhal pointed out in the review Ref.~\cite{terhal2015quantum} that this delay can lead to a backlog problem if we use a decoder that needs to process the whole history of measurement data to extract the outcome of a logical measurement.

To avoid this backlog problem, one can restrict the decoder's memory to a bounded size window (the decoder typically has access to $d$ consecutive rounds of syndrome data) as originally proposed in Ref.~\cite{dennis2002topological}.
This does not completely remove the decoding delay, but this guarantees that it remains constant during a quantum computation.
Then, one could imagine that the decoder is never too slow.
Instead, decoding simply leads to additional idle steps during the implementation of a $T$ gate.
However, these extra steps may make the decoder too slow, causing two potential issues with the decoder's performance:
(i) it may slow down logical gates to the point where there is no more quantum speedup compared to classical algorithms,
(ii) it may increase the depth of logical gates to the point where physical gates are more reliable than logical gates.

Litinski introduced an alternative $T$ state injection circuit~(Fig.17(b) of \cite{litinski2019game}), removing this slowdown at the price of additional ancilla qubits.
This tradeoff can be advantageous in some settings.  
However, it still has a cost and problems (i) and (ii) above remain; after all additional qubits come with more potential fault locations which increases the logical error rate per logical gate, similar to additional time steps.

To summarize, if the decoding delay leads to a significant increase in the space or time cost of a logical gate, the logical error rate per logical gate will go up and one may have to increase the code distance to achieve the target noise rate.
If the decoding delay is too large, it could lead to a negative feedback loop: an increase in code distance comes with a larger decoding delay which leads to an increase code distance etc., making it impossible to achieve the targeted logical error rate.

To avoid reaching the maximum allocated time for the decoder, the decoder can be stopped prematurely once a stopping time is reached.
This reduces the maximum runtime of the decoder at the price of an increased failure rate for the decoder due to timeout failures.
This strategy was proposed in Ref.~\cite{das2022afs} where the stopping time of the decoder was selected in such a way that the timeout failure probability is equal to the failure rate of the uninterrupted decoder.

One may be able to achieve a better performance by selecting more carefully the stopping time of the decoder, which raises the following question:

\bigskip
{\it 
Q3. What is the optimal stopping time for a decoder?
}
\bigskip

We provide two answers to this question motivated by two different scenarios.
In Section~\ref{sec:range}, we consider a surface code with fixed distance $d$ and a decoder, and our goal is to select a stopping time for the decoder that maximizes the computational power of this surface code.
For this purpose, we introduce the notion of range of the decoder (formally defined in Section~\ref{sec:range}).
The range of a decoder estimates the $T$-depth of logical circuits that can be implemented reliably with a given code and decoder. 
Here, by reliably, we mean that the outcome distribution of the logical circuit is error free with probability at least $1-\varepsilon$ for some parameter $\varepsilon$. 
We propose a protocol to identify the stopping time that maximizes the range of a decoder.

The optimization of the stopping time based on the range is relevant for selecting a decoder for a fixed code, but it does not allow for choosing between two decoders because one decoder may have a larger range, but with larger runtime per logical operation.
In Section~\ref{sec:spacetime}, we propose a method to identify the most resource-efficient decoder for a fault-tolerant quantum machine capable of performing a given number of reliable logical operations.
The input is the required $T$-depth $n_T$ for logical circuits.
We select the pair (code distance, stopping time) that achieves a range of at least $n_T$ with a minimum spacetime code per logical operation.
We use the spacetime overhead to measure the cost of the decoder, but our approach is easy to generalize to other cost functions. 
We could for instance adjust the importance of space and time by introducing weights associated with the qubit count or the circuit depth in our cost function.

This idea allows us to associate a single number, the spacetime cost, with any decoder or interrupted decoder. 
By comparing the spacetime cost of different decoders, one can select the most suited one for a given quantum application, or for a class of applications that would require a given logical error rate, providing an answer to our original question Q1.
We envision three classes of practical applications, one at several hundreds to thousand logical qubits with about $10^{-8}$ logical error target, the next is around a thousand to a few thousand logical qubits with around $10^{-12}$, and then the next at $10^{-15}$~\cite{beverland2022assessing}.

We illustrate this approach in Section~\ref{sec:choosing} with the comparison of two theoretical models for decoders with a fast approximate decoder and a slower decoder that has a better accuracy. 
The choice between these two decoders is non-trivial and depends on the targeted $T$-depth.

Our work shows that one may benefit from designing a quantum machine equipped with one or multiple decoders with adjustable stopping times.
The compiler could then select between the available decoders and determine the optimal stopping time for the decoder.

Finally, one may wonder how currently available decoders perform.
We illustrate the concepts introduced in this work with the PyMatching decoder which is fast~\cite{higgott2023sparse} and open source~\cite{pymatching2023}, and answer the following question:

\bigskip
{\it 
Q4. Is PyMatching fast enough?
}
\bigskip

We perform extensive simulations of PyMatching on a desktop machine (Intel Xeon CPU E5-2620 v4 @2.1Ghz with 64Gb of memory) to estimate its runtime distribution.
We measure the runtime of PyMatching for $10^9$ randomly generated fault configurations with $d$ rounds of syndrome extraction and noise rate $p=10^{-3}$.
We repeat these simulations for all odd code distances $d$ from 3 to 31.

Our first observation is that PyMatching is fast enough to render the logical qubits better than the physical qubits for all distances except $d=3$, assuming the syndrome extraction circuit is executed in 1$\mu s$.
Indeed, our numerical results (Fig.~\ref{fig:pymatching_range}) show that the range achievable for surface codes equipped with the PyMatching decoder is larger than the range of physical qubits, except for $d=3$. 
Distance-3 surface codes are not problematic because they can be decoded with a lookup table decoder~\cite{tomita2014low, das2022lilliput}.

However, we find the performance of PyMatching for large code distances is limited.
The range of PyMatching remains bounded below $10^5$ for all code distances and we observe a gap of up to 10 orders of magnitude with the range obtained, assuming no decoding delay (see~\ref{fig:pymatching_range}).

The remainder of this paper is organized as follows.
Section~\ref{sec:computation_model} briefly reviews the surface code.
Our assumptions about the decoder and the delay induced on logical gates are discussed in Section~\ref{sec:decoding_delay}.
The notion of range of a decoder is introduced in Section~\ref{sec:range} where it is used to determine the stopping time maximizing the number of reliable logical gates that can be implemented with a fixed code.
In Section~\ref{sec:spacetime}, we describe a protocol for selecting the code distance and the stopping time that minimizes the spacetime cost of the decoder.
We illustrate how this cost function can be used for choosing between two decoders in Section~\ref{sec:choosing}.
We conclude with potential extensions of this work in Section~\ref{sec:conclusion}.

\section{Computation model}
\label{sec:computation_model}

This background section reviews the surface code and its logical operations.

\medskip
\noindent
{\bf Surface code --}
We consider a single logical qubit encoded with a distance-$d$ surface code, that is a grid of $d \times d$ data qubits.
The surface code was originally introduced by Kitaev on a closed manifold~\cite{kitaev2003fault} and a planar version was proposed in~\cite{bravyi1998quantum, freedman1998projective}.
We consider the rotated version of the planar surface code which is more qubit efficient~\cite{bombin2007optimal}.
To avoid the accumulation of faults, we must constantly run a syndrome extraction circuit which performs local measurement in the grid of qubits using additional ancilla qubits placed in the center of each square plaquette.
In this work, we assume the syndrome extraction is implemented in a gate-based model, and can be implemented in about 1 microsecond with 1 round of preparation, 4 rounds of CNOT gates and 1 round of measurement~\cite{fowler2012surface}.
For other qubit designs, it may be a sequence of measurements~\cite{chao2020optimization, gidney2022pair}. 
The runtime of the syndrome extraction circuit depends on the gate and measurement times and may vary significantly.
The most relevant time scale for our question is not the decoding time itself but how it compares with the syndrome extraction time.
Therefore, we take the runtime of the syndrome extraction circuit as a unit.
We refer to a {\em SEC cycle} as one execution of the syndrome extraction.
The time it takes for one SEC cycle is denoted by $\tSEC$ (in seconds).
In the case of superconducting qubits, a standard assumption is $\tSEC = 10^{-6}$~\cite{gidney2021factor}.
During a SEC cycle, $d^2 - 1$ syndrome bits are extracted (one per surface code plaquette).
These syndrome values are used by the decoder to identify and correct errors.

\begin{figure}
    \centering
    \includegraphics{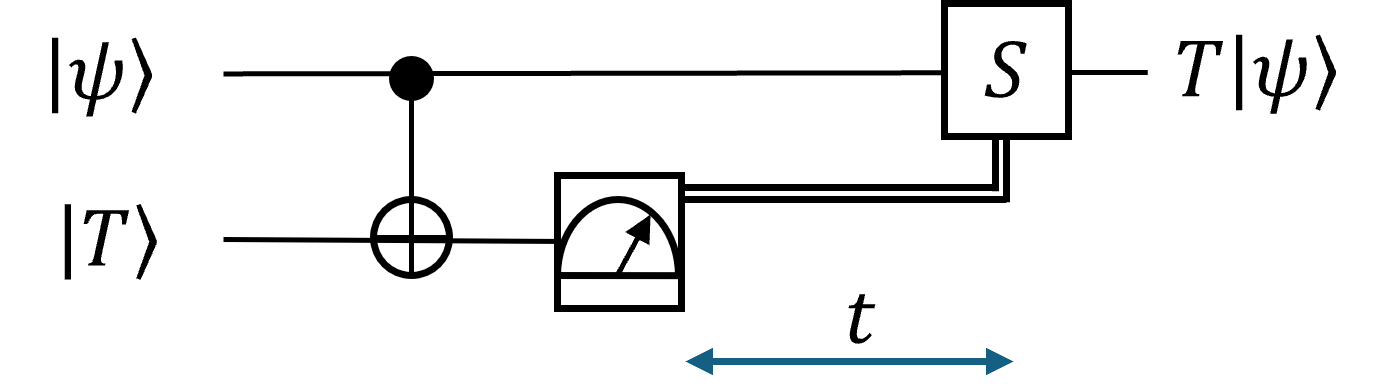}
    \caption{State injection circuit for the implementation of a $T$ gate. Assume that we can prepare a $T$ state $\ket T = \frac{1}{\sqrt{2}}(\ket 0 + e^{i\pi/4} \ket 1)$ on the bottom qubit. Then, we can apply a $T$ gate on the top qubit using only Clifford operations. The gate $S$ is applied conditionally on the outcome of the measurement of the bottom qubit being non-trivial and the delay $t$ between the measurement and the $S$ gate is the time required to extract this measurement outcome.
    When executing this circuit with encoded qubits, the measurement outcome is extracted by the decoder and the delay $t$ is given by the decoding time.}
    \label{fig:Tgate}
\end{figure}

\medskip
\noindent
{\bf Logical operations --}
Operation on logical qubits must be implemented in a fault-tolerant way.
Single-qubit logical Clifford gates $H$ and $S$ are implemented by deformation of the surface code patch using additional qubits; see~\cite{litinski2019game} for a review.
We assume that these gates are implemented in $2d$ SEC-cycles.
One can perform a logical $Z$ measurement in $d$ SEC-cycles ending with the physical measurement of all the $d^2$ qubits of the surface code patch.
These gates are completed with a logical $T$ gate implemented by state injection~\cite{bravyi2005universal} using the circuit shown in Fig.~\ref{fig:Tgate}.
For that, we use a second logical qubit, also encoded with the distance-$d$ surface code, capable of preparing a logical $T$ state.

\medskip
\noindent
{\bf Logical circuit compilation --}
Any single-qubit unitary operation can be written (exactly or approximated) by a circuit of the form~\cite{matsumoto2008representation, giles2013remarks} 
\begin{align} \label{eq:HST_compilation}
HS^{a_1}T HS^{a_2}T \dots HS^{a_{n_T}}T 
\end{align}
made of a sequence of $H$, $S$ and $T$ gates with $n_T$ $T$ gates.
Therein, the parameters $a_i$ take the value 0 or 1.
These sequences can be used to compile any unitary with a minimum $T$ count.
This is a popular choice considered, for instance, in Ref.~\cite{bombin2023modular}.
We use this compilation scheme to implement arbitrary single-qubit logical operations on surface code.
Adding entangling logical operations such as logical CNOT gates or joint measurements, one can achieve a universal set of logical gates.
Our goal is to achieve a low enough logical error rate per logical operation to reliably implement any sequence~\eqref{eq:HST_compilation} for a given number $n_T$ of $T$ gates.

\medskip
\noindent
{\bf Physical error rate --}
The simulations in this paper are implemented with the circuit-level noise model introduced in Ref.~\cite{dennis2002topological}.
Each gate and idle, or waiting, step is followed with probability $p$ by a uniform non-trivial Pauli error acting on the support of the gate and measurement outcomes are flipped with probability $p$.
We use Stim~\cite{gidney2021stim} to generate errors and compute the corresponding syndromes, and to estimate the performance of PyMatching~\cite{pymatching2023}.
We refer to $p$ as the physical error rate.

\medskip
\noindent
{\bf Decoding failure rate --}
The logical error rate per logical operation is traditionally estimated by considering the probability $p_{\fail}(d, p)$ that a logical error appears after correction when running $d$ consecutive SEC cycles with physical error rate is $p$.
This number $p_{\fail}(d, p)$ is called the {\em decoding failure rate}.
To estimate $p_{\fail}(d, p)$ numerically, we perform $d$ rounds of noisy syndrome extraction followed by a round of noiseless syndrome extraction and we check if a logical error occurs.

A popular heuristic estimate of the decoding failure rate of the MWPM decoder based on Fowler's numerical results for the standard CNOT-based syndrome extraction circuit reads~\cite{fowler2012surface}
\begin{align} \label{eq:Plog_heuristic}
p_{\fail}(d, p) = 0.1 \left( 100p \right)^{\frac{d+1}{2}} \cdot
\end{align}
This heuristic is valid for odd distances $d=3,5, \dots$ in the below-threshold regime, {\em i.e.} $p < 10^{-2}$.

Keep in mind that a single logical operation may take more than $d$ SEC cycles.
A rough estimate of the logical error rate of a logical operation implemented in $sd$ SEC cycles for an integer $s$ is $s p_{\fail}(d, p)$.
The logical error rate per logical operation is sometimes used as a metric to estimate the quality of logical qubits but it may be confusing because different logical operations have different logical error rates. 
Moreover, the separation between two consecutive logical operations is not necessarily clearly defined.
To avoid this issue we focus on the probability of an error in the output distribution of the logical circuit that is defined next.

\medskip
\noindent
{\bf Logical circuit error rate --}
Circuit noise affects the outcome distribution of the logical circuit executed. Our goal is to ensure that the code and the decoder are good enough to produce samples from the correct distribution with high probability.
We say that the {\em logical circuit error rate} is below $\varepsilon$ if the error corrected circuit produces a sample from the same distribution as the noiseless circuit with probability at least $1-\varepsilon$.
One can reduce $\varepsilon$ by increasing the code distance or by improving the decoder or both.

\section{Impact of the decoding delay}
\label{sec:decoding_delay}

Here, we introduce notations for the decoding runtime distribution, and we explain how the decoding time translates into a slowdown of the logical $T$ gates, that is a slowdown of the logical clock rate.

\begin{figure}
    \centering
    \includegraphics[scale=.6]{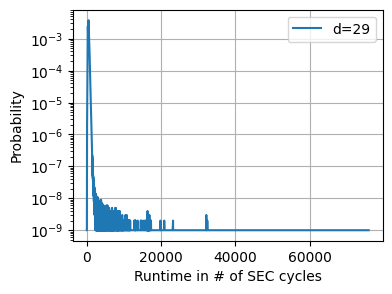}
    \hspace{1cm}
    \includegraphics[scale=.6]{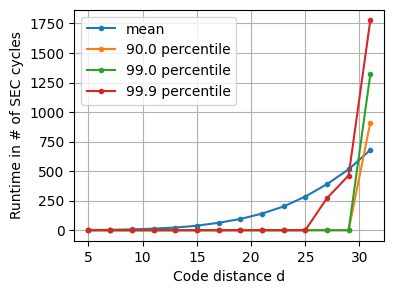}
    
    \hspace{1cm} (a) \hspace{5cm} (b)
    \caption{
    (a) Runtime distribution of PyMatching estimated using $10^9$ decoding trials for a distance-29 surface code with $d$ rounds of syndrome extraction on a desktop computer with an Intel Xeon CPU E5-2620 v4 @2.1Ghz processor with 64Gb of memory.
    We use Stim~\cite{gidney2021stim} to sample circuit faults and to compute their syndromes.
    The runtimes include only the decoding time and do not account for the faults and syndrome generation.
    We assume a SEC cycle time of $\tSEC = 1 \mu s$. The average runtime is around 515$\mu s$ and the corrected standard deviation is around 146$\mu s$.
    (b) Mean runtime and percentiles of the runtime distribution of PyMatching as a function of the code distance. The 99 percentile is the runtime $t$ such that $99\%$ of the samples have runtime $\leq t$.
    We use $10^9$ samples for each code distance. 
    }
    \label{fig:pymatching_runtime_plots}
\end{figure}

\medskip
\noindent
{\bf Decoding time --}
The {\em decoding time} is a random variable that depends on the fault configuration which occurs.
Denote by $\Prob_{\dec}(t)$ the probability that the decoder runs in $t$ seconds.
The decoding time distribution $\Prob_{\dec}$ depends on the code distance $d$ and physical error rate $p$.
The {\em maximum decoding time}, denoted $\tmax$, is the largest value $t$ that  occurs with non-zero probability.
In Figure~\ref{fig:pymatching_runtime_plots}, we plot the runtime distribution of PyMatching for a distance-29 surface code.
We observe that the average runtime of Pymatching is linear in the volume of the syndrome extraction circuit. 

\medskip
\noindent
{\bf Sliding window decoding --}
We consider a sliding window decoder as proposed in Ref.~\cite{dennis2002topological}.
The decoder takes as an input $d$ consecutive rounds of syndrome data, say rounds $r+1, r+2 ..., r+d$ and it returns a Pauli correction on the $n$ data qubits. 
The goal of this correction is to cancel the effect of the faults occurring during the first half of the decoding window, that is the first $t=(d+1)/2$ rounds of this window.
Then, we move the window forward by $t$ rounds and we apply the decoder to the rounds $r+t+1, r+t+2,..., r+d$.

\medskip
\noindent
{\bf Decoding delay --}
In addition to the syndrome data, the decoder also requires as an input the correction estimated for the previous window.
As a result, one needs to wait for the result of the decoding of the previous window.
This is not an issue for a circuit that contains only Clifford operations because the correction can be applied later.
It suffices to propagate it through the circuit and because the circuit is Clifford this correction remains a Pauli correction and can be applied later to qubits.
Decoders capable of decoding multiple windows in parallel were proposed recently~\cite{skoric2022parallel, tan2022scalable, bombin2023modular}.
The situation changes as soon as we hit a non-Clifford gate; a $T$ gate in our case; during which the application of the conditional $S$ gate requires the logical outcome of the measurement of the ancilla qubit which is extracted by the decoder; see Figure~\ref{fig:Tgate}.
This induces a delay of $\lceil \tmax / \tSEC \rceil$ SEC cycles for each $T$ gate.
The {\em SEC-depth}, denoted $\Delta(n_T, d, p, \tmax)$, that is the total number of SEC-cycles required to execute a circuit of the form~\eqref{eq:HST_compilation} is 
\begin{align} \label{eq:SEC_depth}
\Delta(n_T, d, p, \tmax) 
= n_T \left( 7d + \left\lceil \frac{\tmax}{\tSEC} \right\rceil\right) 
\end{align}
where $d$ is the code distance, $p$ is the physical error rate and $n_T$ is the number of $T$ gates.
Therein, the term $7d$ accounts for the implementation of $H$, $S$ and the conditional $S$ gate and the logical measurement.

\medskip
\noindent
{\bf Required distance --}
We are given qubits with a physical error rate $p$ and we want to be able to reliably implement any sequence~\eqref{eq:HST_compilation} with up to $n_T$ $T$ gates. 
Given $p$ and $n_T$, define the {\em required distance} $d$ to be the smallest odd integer $d \geq 3$ such that 
\begin{align} \label{eq:required_distance}
\frac{\Delta(n_T, d, p, \tmax)}{d} \cdot p_{\fail}(d, p)  \leq \varepsilon \cdot
\end{align}
In what follows, the required distance is denoted $\delta(n_T, p, \tmax)$.
The left-hand side is used as a proxy to the logical circuit error rate.
It is only a rough approximation because it does not account for the details of the implementation of the logical gates.
We could refine our estimate of the required distance in Eq.~\eqref{eq:required_distance} by plugging in a more accurate estimate of the probability of a logical error during a logical $S$ gate or a logical $T$ gate or using a better upper bound on the probability of a logical error during a large depth circuit.

The purpose of Eq.~\eqref{eq:required_distance} is to select the smallest distance $d$ ensuring a low logical circuit error rate $\varepsilon$. 
We use $\varepsilon=0.5$ in our simulations. 
One can replace this value by any target $\varepsilon \in [0,1]$ that is appropriate for the application we care to investigate.
For an application with $n$ logical qubits, we may consider replacing the value $\varepsilon=0.5$ by $\varepsilon = 0.5 / n$.
We could also consider the impact of the layout algorithm with restricted connectivity~\cite{litinski2019game, beverland2022surface}.
For simplicity, we stick with a single logical qubit model in the present work.
One could also optimize the value of $\varepsilon$ as a function of the active volume of a quantum algorithm we wish to execute~\cite{litinski2022active}.

In Eq.~\eqref{eq:required_distance}, we assume that we use a surface code with distance at least $3$. 
If $n_T$ is small, we may achieve a sufficiently low noise rate for a circuit with $T$-depth $n_T$ without any encoding.
Namely, if $n_T < \varepsilon / (3p)$, we can implement any sequence \eqref{eq:HST_compilation} with up to $n_T$ $T$ gates with the guarantee that an error occurs with probability smaller than $\varepsilon$ on the entire circuit without any encoding.

\section{Stopping time maximizing the range of a decoder}
\label{sec:range}

In this section, we propose a protocol to identify the stopping time maximizing the $T$-depth of logical circuits that can be implemented with low noise rate on their outcome.
This stopping time can be estimated based on the runtime distribution of the decoder which we can extract from a Monte-Carlo simulation.
We apply our protocol to identify the best stopping time for PyMatching in this context.

\medskip
\noindent
{\bf Interrupted decoder --}
Consider the decoder interrupted after $M$ seconds.
If the initial decoding runtime distribution is $\Prob_{\dec}$, the distribution with this stopping condition is
\begin{align} \label{eq:interupted_decoder_runtime_distribution}
\Prob_{\dec}^{(M)}(t) = 
\begin{cases}
0 \text{ if } $t > M$,\\
\frac{\Prob_{\dec}(t)}{\Prob_{\dec}(t \leq M)}\cdot
\end{cases}
\end{align}
There are two sources of failure for this interrupted decoder.
We say that a {\em decoding failure} occurs if the decoder returns an incorrect correction.
There is a {\em timeout failure} if the decoder does not terminate fast enough (in less than $M$ seconds).
The decoding failure rate of the interrupted decoder, denoted $p_{\fail}^{(M)}(d, p)$, accounts for both types of failures.
It satisfies
\begin{align} \label{eq:interrupted_decoder_pL_bounds}
\max\left( p_{\fail}(d, p), \Prob_{\dec}(t > M) \right) 
\leq p_{\fail}^{(M)}(d, p) 
\leq p_{\fail}(d, p) + \Prob_{\dec}(t > M)
\end{align}
where $p_{\fail}(d, p)$ is the decoding failure rate of the decoder without stopping condition.
In numerical simulations, one can use the upper bound $p_{\fail}(d, p) + \Prob_{\dec}(t > M)$ as an approximation of $p_{\fail}^{(M)}(d, p)$.
This approximation is easy to estimate without further Monte-Carlo simulations and it is correct up to a $2\times$ multiplicative factor because
\begin{align}
\max\left( p_{\fail}(d, p), \Prob_{\dec}(t > M) \right) 
\geq
\frac{p_{\fail}(d, p) + \Prob_{\dec}(t > M)}{2} \cdot
\end{align}

\medskip
\noindent
{\bf Range of a decoder --}
Our goal is to squeeze out as many high-quality logical gates as possible from qubits encoded with a distance-$d$ surface code and equipped with our decoder.
To measure the performance of a decoder as a function of its stopping time $M$, we use the maximum length $n_T$ of a sequence of the form \eqref{eq:HST_compilation} that can be implemented reliably in the sense of Eq.~\eqref{eq:required_distance}.
Formally, define the {\em range of a decoder} as
\begin{align} \label{eq:range}
\range(d, p, M) = \max \left\{ n_T = 0,1,\dots | \frac{\Delta(n_T, d, p, M) \cdot p_{\fail}^{(M)}(d, p)}{d} \leq \varepsilon \right\} \cdot
\end{align}
Varying $M$, we can maximize the range of the decoder, increasing the depth of sequences of logical gates that can be implemented with a logical circuit error rate below $0.5$.
We refer to the stopping time $M$ maximizing the range of the decoder as the {\em range-optimized stopping time} of the decoder\footnote{This number depends on the noise rate $\varepsilon = 0.5$ required for the outcome distribution of the logical circuit.}.

Later in this paper, we optimize the stopping time in a different way.
We allow to vary both the code distance and the stopping time, and we minimize the spacetime overhead to implement a fixed number of logical $T$ gates.

\medskip
\noindent
{\bf Bounds on the range --}
Quantum error correction is useful only if it boosts the performance of our quantum hardware.
Without any encoding, one can achieve a range 
\begin{align} \label{eq:range_no_encoding}
\left\lfloor
\frac{\varepsilon}{3p}
\right\rfloor
\end{align}
where $\varepsilon$ is the allocated probability for an error in the whole circuit, and $p$ is the physical error rate.
The constant $3$ comes from the compilation in $HST$ sequences.
To be useful, a decoder must achieve a range larger than Eq.~\eqref{eq:range_no_encoding}, which is $166$ for $p=10^{-3}$ and $\varepsilon = 0.5$.

An upper bound on the range of a decoder is obtained by assuming that the decoder is instantaneous, that is by setting $\tmax = 0$ in Eq.~\eqref{eq:SEC_depth}.

\begin{figure}
    \centering
    \includegraphics[scale=.6]{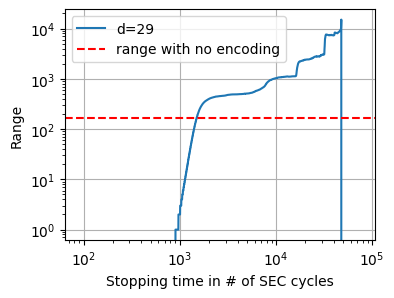}
    \hspace{1cm} 
    \includegraphics[scale=.6]{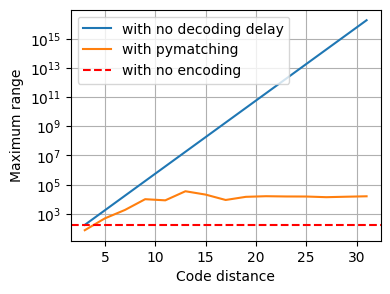}
    
    \hspace{1cm} (a) \hspace{5cm} (b)
    \caption{
    (a) Range of the distance-29 surface code with the PyMatching decoder. This range is computed based on the runtime distribution shown in Figure~\ref{fig:pymatching_runtime_plots}.
    The red dashed line represents the range achievable without encoding given in Eq.~\eqref{eq:range_no_encoding}. It is $166$ for $p=10^{-3}$ and $\varepsilon = 0.5$. 
    (b) Range of the surface code equipped with the PyMatching decoder as a function of the code distance and comparison with the same decoder assuming no decoding delay.
    The runtime of PyMatching has a significant impact on the range of the decoder which drops by more than 10 orders of magnitude at distance 29.
    Due to the decoding delay, the range remains mostly flat beyond distance 13.
    }
    \label{fig:pymatching_range}
\end{figure}

\medskip
\noindent
{\bf Example: Maximum range of PyMatching --}
As an example, we analyze the performance of PyMatching~\cite{higgott2021pymatching, higgott2023sparse, pymatching2023}, which is a fast implementation of the MWPM decoder for surface codes.
Our results are plotted in Fig.~\ref{fig:pymatching_range}.
We use a desktop computer with an Intel Xeon CPU E5-2620 v4 @2.1Ghz with 64Gb of memory for these simulations and Stim~\cite{gidney2021stim} is used to generate the surface code circuit and the faults.
For each code and for each odd distance $d$ from 3 to 31, we run $10^9$ decoding trials of PyMatching and we measure the decoder runtime (excluding the time for error and syndrome generation). 
Each runtime sampled is the decoding time for the correction of a set of circuit faults for $d$ rounds of syndrome extraction with a distance-$d$ surface code, with noise rate $p = 10^{-3}$.
This provides an estimate of the runtime distribution of PyMatching as we can see for distance 29 in Fig.~\ref{fig:pymatching_runtime_plots}.
The runtime is measured in SEC cycles with the assumption that a SEC cycle takes $\tSEC = 10^{-6} s$ (in other words, the runtime is measured in $\mu s$).
We also keep track of the decoding failure, so that for each stopping time $M$, we can count the number of decoding failures of the interrupted decoder (including decoding failures and timeout failures).
To keep only the results that are statistically significant, we only consider the stopping times for which at least 20 failures are observed.
This lets us estimate $p_{\fail}^{(M)}(d, p)$ for all $M$ (here $p=10^{-3}$) and using Eq.~\eqref{eq:range}, we estimate the corresponding range as 
\begin{align} \label{eq:pymatching_range_estimate}
\left\lfloor
\frac{\varepsilon d}{p_{\fail}^{(M)}(d, p)(7d + M)} 
\right\rfloor
\end{align}
where $\varepsilon = 0.5$.

In Fig.~\ref{fig:pymatching_range}, we compare PyMatching with an instantaneous decoder that induces no delay during the $T$ gate.
We estimate the range of the instantaneous decoder using Eq.~\eqref{eq:pymatching_range_estimate} with 
$M = 0$ and 
$
p_{\fail}^{(M)}(d, p) = 0.04 (0.1)^{(d+1)/2}
$.
This estimate for the decoding failure rate of PyMatching is extracted from our simulation.

We observe in Fig.~\ref{fig:pymatching_range} that PyMatching decoder is fast enough to make the surface code useful in some regime because it outperforms the unencoded range.
However, there is significant room for a faster decoder for large code distances.
Indeed, the maximum range drops by more than 10 orders of magnitude at distance 29 compared to the range obtained by assuming that the decoder is instantaneous.

\begin{figure}
    \centering
    \includegraphics[scale=.6]{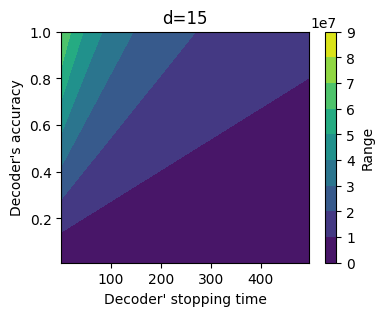}
    \hspace{1cm} 
    \includegraphics[scale=.6]{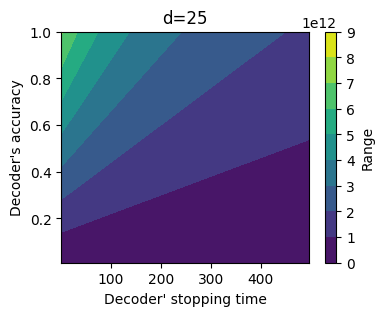}

    \hspace{1cm} (a) \hspace{5cm} (b)
    \caption{
    Range of the decoder as a function of its accuracy and its stopping time in number of SEC cycles for distance 15 in (a) and 25 in (b).
    The first level (separating the two darkest regions) corresponds to the tradeoff required to achieve a range of $10^7$ for d=15 (and $10^{12}$ for $d=25$), that is to allow for the reliable implementation of a logical circuit with $10^7$ $T$ gates.
    }
    \label{fig:decoder_req}
\end{figure}

\medskip
\noindent
{\bf Decoder requirements for large depth circuits --}
A faster decoder is needed to preserve most of the performance of surface codes with large distances.
In Fig.~\ref{fig:decoder_req}, we examine the tradeoff between the decoder's accuracy and its stopping time.
We say that a surface code decoder has {\em accuracy} $\alpha \in [0, 1]$ if it achieves a failure rate equal to 
$
p_{\fail}(d, p) / \alpha
$,
where $p_{\fail}(d, p)$ is the standard heuristic of Eq.~\eqref{eq:Plog_heuristic}.
We plot the range of a decoder as a function of its accuracy $\alpha$, and its stopping time $M$.
Similarly to Eq.~\eqref{eq:pymatching_range_estimate}, this range is computed as 
\begin{align}
\frac{\varepsilon d \alpha}{p_{\fail}(d, p)(7d + M)}    
\end{align}
for the distance-$d$ surface code.

Consider for example the plot of the distance 15 range.
We observe that, if one can design a decoder whose stopping time is only a few SEC cycles, it only needs to achieve an accuracy of $0.2$ to reach a range of $10^7$.
If the decoder' stopping time is larger, say 500, it must be reach a $0.8$ accuracy to obtain the same range.
However, this may make the logical operation very slow. Indeed, the time for a logical $HST$ sequence would go from $7d = 105$ without decoding delay, to $7d + M = 605$ with a stopping time equal to 500 SEC cycles.
We observe a similar phenomenon for other code distances.

\section{Stopping time minimizing the spacetime cost of a decoder}
\label{sec:spacetime}

In this section, we consider a decoder for the family of surface codes.
We vary both the code distance $d$ and the stopping time $M$ and our goal is to select the pair $(d, M)$ that allows for a sufficiently large number of reliable logical gates and that minimizes the spacetime cost per logical gate.

\medskip
\noindent
{\bf Spacetime cost --}
We are given qubits with a noise rate $p$ and a SEC cycle time $\tSEC$ and our goal is to design a quantum computer capable of performing $n_T$ logical $T$ gates reliably.
Again by reliably, we mean that the logical circuit error rate is at most $\varepsilon$.
Define the {\em spacetime cost} of the decoder to be
\begin{align} \label{eq:encoded_volume}
\cost(p, n_T, d, M) = 
\begin{cases}
+\infty \text{ if } \range(d, p, M) < n_T,\\
2d^2 \Delta(n_T, d, p, M) \text{ otherwise}
\end{cases}
\end{align}
where $M$ is the stopping time of the decoder.
By {\em minimum spacetime cost} of a decoder, we mean the minimum spacetime cost of a decoder for all odd distances and for all stopping time.
We denote the minimum spacetime cost as
\begin{align}
    \mincost(p, n_T) = \min \left\{ \cost(p, n_T, d, M) \ | \ d \text{ odd}, M > 0 \right\} \cdot
\end{align}
This number can be infinite if $p$ is above threshold.
The pair $(d, M)$ that achieves this minimum spacetime cost is denoted 
$d, p = \argmincost(p, n_T)$.

To pick the most suitable decoder for a fault-tolerant quantum computer designed for $n_T$ logical $T$ gates, we select the decoder with the smallest minimum spacetime cost and we pick the corresponding code distance and stopping time.

\begin{figure}
    \centering
    \includegraphics[scale=.6]{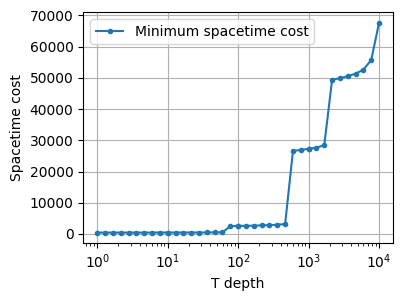}
    \hspace{1cm} 
    \includegraphics[scale=.6]{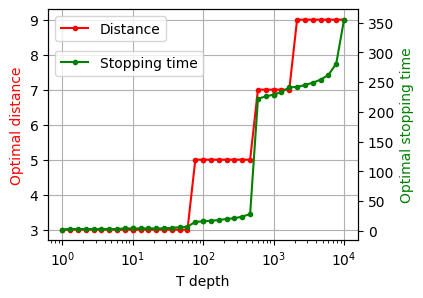}
    
    \hspace{1cm} (a) \hspace{5cm} (b)
    \caption{
    (a) Minimum spacetime cost of PyMatching as a function of the required number of logical $T$ gates $n_T$ for $p=10^{-3}$ and with a SEC cycle time of $1\mu s$.
    (b) Distance and stopping time achieving the minimum spacetime cost as a function of $n_T$.
    In these plots, we only consider logical circuits with a $T$ depth up to $10^4$ because we could not reach $10^5$ reliable logical $T$ gates with this decoder (See Fig.~\ref{fig:pymatching_range}(b)).
    }
    \label{fig:pymatching_min_spacetime_cost}
\end{figure}

\medskip
\noindent
{\bf Example: Spacetime cost of PyMatching --}
In section~\ref{sec:range}, we estimated the range of PyMatching as a function of the code distance $d$ and the stopping time $M$.
Using this information, we can compute the spacetime cost of PyMatching for all values $d$ and $M$ for which we observed at least 20 failures (to guarantee statistical significance) and extract the minimum spacetime cost.
We estimate the minimum spacetime cost for a range of values of $n_T$ between 1 and $10^4$ and our results are presented in Fig.~\ref{fig:pymatching_min_spacetime_cost}.
We stopped at $n_T = 10^4$ because the spacetime cost of PyMatching becomes infinite when we reach $10^5$.
This is because the maximum range of all the codes we simulated (up to distance 31) is below $10^5$ as we can see in Fig.~\ref{fig:pymatching_range}.
The jumps in the minimum spacetime cost correspond to the increase in code distance.

\section{Choosing between two decoders}
\label{sec:choosing}

In this section, we illustrate our protocol to select the most suited of two decoders.
Our main goal is to show that the landscape of the spacetime cost function is non-trivial which makes it challenging to identify the most cost-efficient decoder.
We use two theoretical models for the decoder runtimes. 
One decoder has a smaller maximum runtime and the other one has a more favorable decoding failure rate.

We consider two decoders whose runtime distribution is a binomial distribution with parameters $N$ and $Q$.
The runtime of the corresponding decoder is $t$ SEC cycles with probability $\Prob(t) = \binom{N}{t} Q^t (1-Q)^t$, where $t \in \{0,1,\dots, N\}$.
To define such a runtime distribution, we can provide $N$ and $Q$.
Alternatively, it is enough to give the mean runtime, which is equal to $NQ$ and the maximum runtime (equal to $N$).

Suppose that we have qubits with a physical error rate $p = 10^{-3}$ and a SEC cycle time of $1 \mu s$, {\em i.e.} $\tSEC = 10^{-6}$, which is similar to the case of several fast qubit architectures.
The first decoder we consider is a {\em quadratic-time decoder} that achieves the same decoding failure rate $p_{\fail}(d, p)$ as in Eq.~\eqref{eq:Plog_heuristic} with a mean runtime of $p d^3 \mu s$ and a maximum runtime of $d^6 \mu s$.
Our second decoder is a {\em linear-time decoder}. It runs in linear time in the worst case and it is four times faster on average with a mean runtime of $0.25 p d^3 \mu s$ and a maximum runtime of $0.25 d^3 \mu s$.
However, its decoding failure rate is larger, equal to $(4/3) p_{\fail}(d, p)$.

\begin{figure}
    \centering
    \includegraphics[scale=.6]{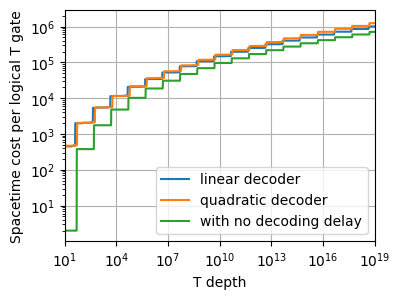}
    \hspace{1cm} 
    \includegraphics[scale=.6]{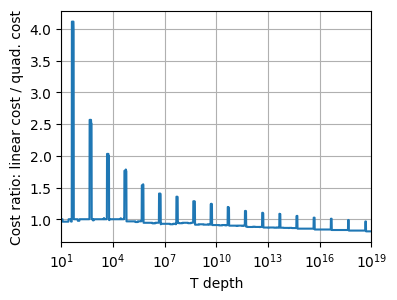}
    
    \hspace{1cm} (a) \hspace{5cm} (b)
    \caption{
    (a) Spacetime cost of the linear-time decoder, the quadratic-time decoder and the instantaneous decoder that leads to no delay during the $T$ gate.
    We assume that decoding failure rate of the instantaneous decoding is given by Eq.~\ref{eq:Plog_heuristic} like for the quadratic-time decoder.
    (b) Ratio between the spacetime cost of the linear-time decoder and the quadratic-time decoder.
    }
    \label{fig:binomial_decoders}
\end{figure}

To select the best decoder for a given fault-tolerant quantum machine built for $n_T$ reliable logical $T$ gates, we compare their spacetime cost.
Our numerical simulation in Fig.~\ref{fig:binomial_decoders} shows that the choice depends on $n_T$ in a subtle way. 
Selecting the decoder with minimum cost is non-trivial.
The cost of the linear-time decoder can be up to 4 times larger than the cost of the quadratic-time decoder for small $n_T$.
However, for large $n_T$ the linear-time decoder becomes cheaper with a cost that is about $80\%$ of the cost of the quadratic-time decoder.

\section{Conclusion}
\label{sec:conclusion}

In this work, we propose a strategy to select a decoder for a given fault-tolerant quantum computer architecture, including how to optimize the stopping time of a decoder.
The choice of the decoder and the optimization of its stopping time depends on the physical qubit noise rate of the underlying quantum hardware, the speed of gates and measurements, and also the compilation algorithm used to compile logical gates into fault-tolerant gates and the magic state injection circuit.
Here we assumed a simple, well-studied parameter setting; future work could consider the impact of different choices for these parameters on the choice of decoder.
Further work could also consider more realistic models by replacing the transversal logical CNOT gate assumed here by a more realistic CNOT implementation, or by varying the number of logical qubits and the reliability parameter $\varepsilon$.
In practice the logical qubit connectivity may affect the cost of logical operations~\cite{litinski2019game, beverland2022surface} and the decoder optimization.
One could also investigate the impact of the decoding delay inside the distillation circuit, which is ignored in this work and was previously considered in~\cite{bombin2023modular}.
Ultimately, one could include the impact of the decoding delay in resource estimation tools~\cite{beverland2022assessing}.

Here, we measure the decoding cost through the spacetime cost of logical operations (accounting for decoding delays) and we ignore some architecture constraints such as memory and energy~\cite{das2022afs, holmes2020nisq+, ueno2021qecool}.
One may consider refined metrics for the cost of a decoder to capture decoding hardware requirements such as memory cost, energy or footprint. 
The bandwidth requirements can also be significant for running large-scale quantum algorithms~\cite{delfosse2020hierarchical, smith2023local}.

We use our framework to analyze the cost of PyMatching~\cite{pymatching2023} running on a Windows desktop machine. 
Our results show that this implementation can be fast enough for the execution of thousands of logical $T$ gates with small surface codes, which is a significant result for a decoding software which was originally designed as a simulation tool.
One could get better results by running this software on a more powerful machine or by further optimizing the code.
One may also be able to speed-up this algorithm using a pre-decoder as proposed in~\cite{delfosse2020hierarchical} with the lazy decoder or in~\cite{caune2023belief} using BP as a pre-decoder.

We observe a large variability in the decoder runtime that may be due to the interference of the operating system.
Running the decoder on a dedicated machine may help to prevent large variations of the runtime.
A FPGA or an ASIC implementation could solve this issue as in~\cite{liyanage2023scalable, barber2023real}.

Overall, one could use our protocol to select between different decoding algorithms for a given clasical hardware implementation, say a desktop machine or an FPGA.
More broadly, it could be used to compare two different decoding systems which include a decoding algorithm and the hardware it runs on.

Ultimately, in designing a quantum computer to scale to practical quantum advantage, one must co-design the hardware and software, together.  
To achieve a quantum supercomputer will require carefully considering the underlying physical qubit speeds and fidelities, and designing a decoder to achieve the desired specifications and performance of that quantum machine. 
For superconducting qubits with a SEC cycle time of $1 \mu s$, one can decode a machine capable of implementing $10^7$ reliable logical $T$ gates with a decoder that achieves $50\%$ of the accuracy of the MWPM decoder with a maximum runtime of $250 \mu s$ (see Fig.~\ref{fig:decoder_req}(a)).

\section{Acknowledgments}

The authors would like to thank Marcus Silva, Michael Beverland and Rui Chao for insightful discussions during the preparation of this manuscript.


\end{document}